# Traffic signal control optimization under severe incident conditions using Genetic Algorithm

Tuo Mao [12*], Adriana-Simona Mihăiţă [12], Chen Cai [1]

[1]Advanced Data Analytics in Transport, DATA61 | CSIRO, Level 5, 13 Garden St, Eveleigh NSW 2015, Sydney, Australia
[2]University of Technology Sydney, 15 Broadway, Ultimo NSW 2007, Sydney, Australia
*tuo.mao@uts.edu.au

**Abstract:** Traffic control optimization is a challenging task for various traffic centres in the world and majority of approaches focus only on applying adaptive methods under normal (recurrent) traffic conditions. But optimizing the control plans when severe incidents occur still remains a hard topic to address, especially if a high number of lanes or entire intersections are affected. This paper aims at tackling this problem and presents a novel methodology for optimizing the traffic signal timings in signalized urban intersections, under non-recurrent traffic incidents. The approach relies on deploying genetic algorithms (GA) by considering the phase durations as decision variables and the objective function to minimize as the total travel time in the network. Firstly, we develop the GA algorithm on a signalized testbed network under recurrent traffic conditions, with the purpose of fine-tuning the algorithm for crossover, mutation, fitness calculation, and obtain the optimal phase durations. Secondly, we apply the optimal signal timings previously found under severe incidents affecting the traffic flow in the network but without any further optimization. Lastly, we further apply the GA optimization under incident conditions and show that our approach improved the total travel time by almost 40.76%.

## 1. Introduction

Traffic incident management plays an important role for all transportation agencies because of its impact on safety and traffic control operations. To deal with stochastic incidents, various traffic management centres (TMCs) develop policies and response plan strategies in order to minimize the clearance time. Traffic information and control systems (TIMs) are key components in securing an instant response time since they are centralized and can easily alert the incident to TMCs. The typical response plan applied by many TMCs in case of an emergency/accident is to activate a range of variable message signs, close lanes and force turnings, without having an adaptive control method for signal groups in the affected intersections; most of the time this is a manual process which requires waiting for the incident to be cleared-off until the adaptive control plans are re-activated.

Traffic congestion is classified into two types: recurrent congestion (RC) which can appear due to daily travel patterns and non-recurrent congestion (NRC) which can be caused by unexpected events such as accidents/breakdowns/etc. [1-3]. The most problematic incidents can occur at random locations, at various moments in time and do not ever repeat themselves [1]. It is a big challenge to model and handle the network optimization under these non-recurrent incidents because of its uncertainty of occurrence in both time and space. To the best of our knowledge, there are not many works which focus on traffic signal control optimization under severe incident conditions due to the high variability of traffic conditions and incident incertitude.

This research tries to address this problem and focuses on modelling a new traffic management solution to ease the impact of non-recurrent traffic incidents, by making use of the power of Genetic Algorithm (GAs) and a new green split definition. In this paper, we present an efficient GA which can be applied as a tool for a fast traffic incident response and optimization of the traffic signal control plan. Section 2 presents the current literature review for traffic incident response and signal control modelling, Section 3 focuses on the methodology of our work while Section 4 presents the optimization process and algorithm definition. In section 5 and 6, a case study and its results are discussed. In section 7, conclusions are drawn according to the case study. Overall, the main contributions of this paper are the following:

1. Propose a new traffic signal control optimization method using GAs with the purpose of minimizing the total travel time in urban networks affected by incidents;
2. Employ the phase green splits in our optimisation problem to be decision variables while the traditional methods use the link green split;'
3. Couple the GA fitness function together with a simulation framework consisting of a static assignment followed by a microscopic stochastic route choice simulation in Aimsun.
4. Showcase the dramatic travel time reduction before and after deploying the GA for signal optimisation of an affected road network.

## 2. Literature review

### 2.1 Traffic incident response related work

Current research on traffic incident response is majorly focused on incident response planning and decision making. Ban, et al. [4] recently developed a decision-making tool to determine whether or not to activate the control system when an incident is reported to the traffic management centre by using regression models and support vector machines to quantify the performance of traffic signals in the network. In the official traffic incident management handbook for U.S.A [5], two major tasks are suggested for traffic control such as: actively managing the traffic control devices in the incident affected area and designating alternative routes, but no further details about how to adjust the traffic control are discussed. These manual techniques highly rely on the practical experience of the operators. Therefore, developing an automatic control process in traffic modelling could improve the current traffic management and provide better traffic control suggestions to the operators.

Mehran [6] summarized several TIMs deployments in Asia, Europe, and North America in which the major response to motorists and drivers in case of an accident is to provide information on the current traffic condition, route information and travel time. But the need to minimize the impact of incidents on road traffic is barely focused and investigated. Nitsche, et al. [7] evaluated novel technologies in TIM under different incident scenarios by assessing the discovery time, verification time and initial response time when using a cell transmission model (CTM-v) as a simulation model. Recently, a research about coordinating TIM and Congestion Management (CM) was facilitated by the University of Washington [8] in order to identify the current "as-is" TIM and CM processes, and exploring the desired interventions ("to-be" models); the coordination of TIM and CM was on demand and was regarded as a future exercise in the U.S.A.

In Australia, practices in traffic incident response have been mostly focused on procedures for incident detection, verification, response plan, site management, investigation, clearance, traffic management and traveller information [9]. In this recent report, the microscopic simulation was mentioned as a planning, operation and training tool for helping the TIM systems, but no further detail for any existing and operating microscopic simulation investigation was provided. The report is majorly focused on regulating and refining the traffic response plan for multiple agencies such as Emergency Medical Services (EMS), police and fire stations, but does not provide deep knowledge about the traffic control system. Therefore, there is a true need to improve the current TIM system for handling incidents and adapt automatically the traffic control plans under severe accident condition.

After all, most TIM systems are majorly focused on handling the traffic incident site but barely extended to minimize the impact on the surrounding traffic. To the best of our knowledge, no traffic modelling has been intensively studied and applied to traffic incidents and incident clearance time reduction. On the long-term, this work contributes to our ongoing objective to build a real-time platform for predicting traffic congestion in Sydney, and to analyse the incident impact during peak hours (see our previous works published in [10]-[11]).

*2.2 Traffic signal control modelling*

Current traffic signal control models are refined to deal with mostly recurrent congestions in the network, but they are not so sensitive to the congestion caused by non-recurrent traffic incidents. Severe traffic incidents may strongly influence the traffic signal control and should not be neglected. A well-concluded review published in [12] presented the traffic control modelling for both arterial roads and motorway. In this review, a "store-and-forward model" is introduced to simplify the model-based optimization method by enabling the mathematical description of the traffic flow process without discrete variables; as well it uses the Traffic-response Urban Control (TUC) strategy for calculating the real-time network splits [13]. Ritchie [14] introduced multiple real-time knowledge-based expert systems (KBES) to the advanced traffic management (ATM) system in order to provide suggestions to the control room staff when non-recurrent congestion happened. At that time, the cooperation of artificial intelligence (AI) and ATM were very pioneering and the combination of AI and ATM became a good direction for later research. This conceptual design can be fulfilled now by recent machine learning techniques and big-data processing.

Among various models, GA is popular for its efficiency of optimizing traffic signal controls which was first introduced by Goldberg and Holland [15] in 1988 later applied to traffic signal timing optimization in 1992 in [16]. In 2004, Ceylan and Bell [17] applied stochastic user equilibrium to model the driver's route choice under different signal timings while using GA to optimize the traffic signal timing. It was also concluded that GA is simpler and more efficient than previous heuristic algorithms. GA has been successfully used as well for a multi-objective control plan optimization method for choosing the most effective traffic control plan in [18], but none of the studies applied GA to ease accident affecting the traffic congestion.

Overall, there is still a gap in researching the most efficient and fast response in traffic signal control modelling in order to deal with non-recurrent traffic incidents. Our approach and methodology try to address these problems by deploying an innovative GA modelling while also optimizes the green time splits in intersections affected by incidents, obtaining the minimal travel time. The procedure and description of all the steps are provided in Section 3.

## 3. Methodology

*3.1 Problem formulation*

There are four different steps for creating a traffic incident response: incident identification, verification, response, and clearance. This paper is basically focused on the modelling of traffic management and control after an incident has been confirmed and reported by TMC. The proposed model is going to be applied in the response phase and clearance phase. To simplify the case study, this paper assumes that the incident was previously detected and verified and the duration of the incident clearance was predicted. In addition, the severity of the incident is also reported as an indication of the number of lanes affected.

Last but not least, the incident affected area is determined using previous studies. Recently, Pan, et al. [19] studied the spatial-temporal impact of traffic incidents based on archived data using advanced sensors and came up with the incident impact area and the delay occurrence prediction in a road network. The affected area normally contains all the surrounding network which experiences the congestion caused by the incident and it is generally time-dependent to the reported location of the incident. The problem we are trying to solve is how to optimize the traffic control plan around the incident location, in order to minimize the impact of the incident in terms of vehicle total travel time. Therefore, we use the road network in the affected area which is pre-determined, and we formulate the problem as follows:

Given a road network which has been identified as affected by an accident, we define the following:

- $A$ is the set of links in the network,
- $W$ is the set of origin-destination pairs of the network,
- $R_w$ is the set of routes between origin-destination pair $w \in W$,
- $d_a$ is the queuing delay at link $a \in A$,
- $f_r^w$ is the flow on route $r \in R_w$,
- $v_a$ is the link flow on link $a \in A$,
- $\lambda_a$ is the "link green split" $\lambda_a$ which is determined by traffic signals at the end of the link (the definition will be discussed in the next section),
- $t_a(v_a, \lambda_a)$ is the travel time on link $a \in A$ described as a function of link flow $v_a$ and "link green split" $\lambda_a$,
- $S_a$ is the capacity of link $a \in A$,
- $\sigma_{ar}^w$ is 1 if route $r$ between O-D pair $w$ uses link $a$, and 0 otherwise,
- $D_w$ is the demand between O-D pair $w \in W$,

The target is to minimize the total travel time of the network. The target objective function is as follow:

$$\text{minimize} \sum_{a \in A} \int_0^{v_a} t_a(v_a, \lambda_a) dx \quad (1)$$

Subject to

$$\sum_{w \in W} \sum_{r \in R_w} f_r^w \sigma_{ar}^w = v_a, a \in A \quad (2)$$

$$\sum_{r \in R_w} f_r^w = D_w, w \in W \quad (3)$$

$$v_a \leq \lambda_a S_a, a \in A \quad (4)$$

$$f_r^w \geq 0, r \in R_w, w \in W \quad (5)$$

Equation (2) represents the relation between route flows ($f_r^w$) and link flows ($v_a$). Equation (3) shows the flow conservation between route flows and O-D demands. Equation (4) shows that link flow is limited by the exit capacity, which depends on the link capacity and link green split. Equation (5) indicates that link flows must be no less than zero.

### 3.2 The definition of link green split $\lambda_a$

In this paper, the definition of "link green split" ($\lambda_a$) is the same as the one in the study of Yang and Yagar [20], which is the amount of green time granted for a link (link $a$) in a signalized intersection. As for Smith and Van Vuren [21], green time is divided into: stage green time (or phase green time) and link green time. A phase is defined as a maximal set of compatible approaches in an intersection. Therefore, the stage green time (or phase green time) is the green time of certain stage (or phase) in a cycle in a signalized intersection. The link green time is the green time granted for a link by all the corresponding phases in a cycle of a signalized intersection.

Let $\Lambda_{jk}$ be the proportion of green time for which the $k^{\text{th}}$ phase at junction $j$, therefore we can call $\Lambda_{jk}$ a "phase green split". The allocation of green time to all phases at a junction determines the green time of each link entering that junction, therefore for each link $a$, the "link green split" ($\lambda_a$) is the summation of all those phase green splits ($\Lambda_{jk}$) for which phase $k$ at junction $j$ contain the movement of link $a$, or:

$$\lambda_a = \sum_{\text{stages } S_{jk} \text{ contain link } a} \Lambda_{jk}. \quad (6)$$

To be clear, for each junction $j$, the sum (over k) of "phase green split" $\Lambda_{jk}$ will be 1:

$$\sum_k \Lambda_{jk} = 1. \quad (7)$$

Actually, by defining the Equation (7), we assume that there is no cycle loss time in each cycle of an intersection. In addition, we assume that the amber (yellow) time for each phase is considered as the green time. In conclusion, the $\lambda_a$ in this paper is the "link green split" other than the "phase green split".

### 3.3 Assumptions

In this paper, we assume that the O-D demands are predefined and fixed. We use a traffic assignment model to get the link traffic flows which depend on link cost functions and O-D demands. Therefore, we can get deterministic link flows. In addition, the link travel time function (or cost function) is fixed for all links in the investigated road network which only depends on the link flow and the "link green split". Therefore, the only parameter we try to optimize for each link is the "link green split" $\lambda_a$.

For traffic signals in the network, we assume that each phase of a cycle grants green to fixed movements. The order of phases in a cycle is also fixed. Only the duration of each phase is tuneable. The duration of all phases in all signalized intersections are actually the decision variables for the optimization problem.

## 4. Optimization process

The introduction in "link green split" to our problem leads to an optimization problem for traffic signal timing because of the direct relationship between "link green split" and "phase green split" in Equation (6) and (7). Now the optimization problem can be transformed into the optimization of the traffic signal timing in a road network.

### 4.1 Data input

The specification of the network is required as an input, which consists of:
- O-D configuration: contains the location of origins and destinations,
- O-D demand table: contains the trips between each pair of origin and destinations,
- Network configuration: contains all information about links, nodes, speed limits, road capacity, etc.
- Link detail table: contains link free-flow travel time, link speed limit, link capacity, and number of lanes,
- Traffic signal configuration: signalized node indexes, number of phases, cycle time, signal timings, phase green splits, and the links granted green for each phase.

### 4.2 Genetic algorithm specification

Ideally, we could sample all the possible traffic signal control plans in order to get the optimal traffic signal control plan. As we can see, it is very computationally intensive to sample all possible traffic signal control plans. Let's consider, for example, one signalized intersection which has 4 phases. Each phase has a duration ranging between minimum 3 and maximum 90 seconds, which must be an integer. This means a total of $(90 - 3 + 1)^4 = 59,969,536$ possible traffic control plans. The computational times to test all of the phase combinations to find the optimal solution can be quite intensive just for one intersection, not to mention more complicated road networks with various nodes and complicated connections. Therefore, a GA solution is used to reduce the computational load; the algorithm randomly samples from the total feasibility space of phase combinations and chooses the most representative ones which would minimize the total travel time in the urban network, under recurrent and non-recurrent traffic conditions. The full description and application are provided in the next section.

In our study, we employ a standard GA for traffic signal control optimization which we adapt to our network needs and reported traffic incident. In the following, we detail the parameters and steps we have followed to successfully deploy such model for traffic control plan optimization.

- **Fitness function**: To adapt our problem to GA, the target function in Equation (1) is utilized as the fitness function. As we want to minimize Equation (1) then we want to use the reverse of Equation (1) as our fitness because we maximize the fitness value in GA. Then the fitness value is shown in Equation (8).

$$Fitness = -\sum_{a \in A} \int_0^{v_a} t_a(v_a, \lambda_a) dx \qquad (8)$$

- **The decision variable**: The decision variable is a vector of all phase durations within the network. In order to optimize the target function (Equation (1)), we need to code the decision variables as the chromosome in GA. The coding process is illustrated as follows:

  Decision variables $\psi$ (array of arrays) =
  [ [$p11, p12, p13, p14$] , [$p21, p22, p23, p24$] ,
  ... , [$pn1, pn2, pn3, pn4$]]

  Chromosome (array) =
  [ $p11, p12, p13, p14, p21, p22, p23, p24,$
  ... , $pn1, pn2, pn3, pn4$]

  Where $p_{uv}$ means the phase duration of intersection $u$ phase $v$ and $n$ is the total number of signalized intersections. As we can see, the chromosome in GA is the same as the decision variable with less groupings.

- **The GA solution for traffic signal optimization**: is shown in Fig. 1 and contains various modules such as "check stop", "tournament", "crossover" and "mutation" which are also adapted to our application.

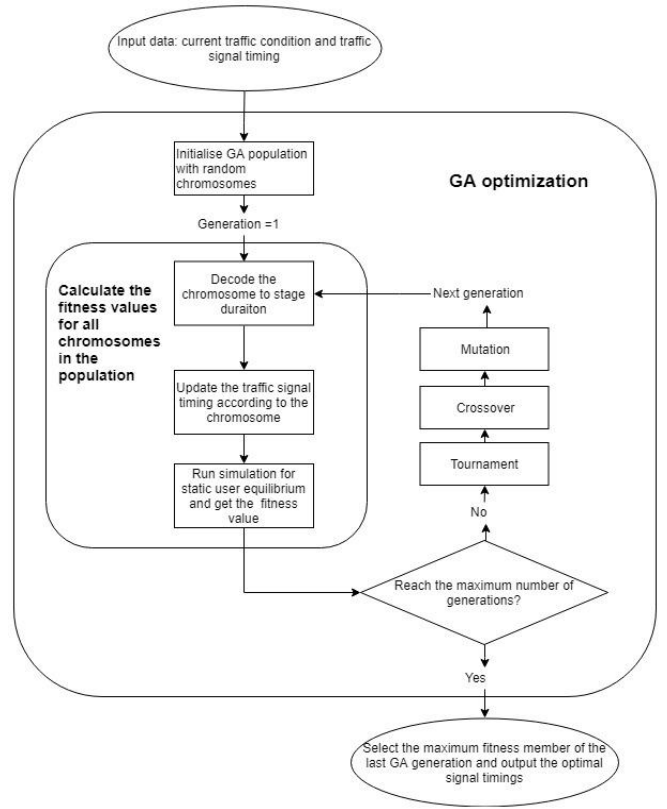

*Fig. 1.* GA optimization process

A detailed description of these modules is the following:

1. **Prepare input data**: Within GA there are several parameters that need to be determined in order to get a fast convergence and a short computation time. We first use the current traffic condition and traffic signal timing but also fix the population size, maximum number of generations, probability of crossover, and probability of mutation.
2. **Initialization**: initialize the GA population with random chromosomes of the dataset.
3. **Fitness function calculation**: for each individual we calculate the fitness function by decoding the chromosomes to phase durations, updating the traffic signal timing according to the chromosome and running a simulation model of the network for static user equilibrium. We used the AIMSUN as our simulation tool to generate the fitness function. Within this function, we called AIMSUN to firstly assign the pre-set OD demand to the network and then run a microscopic stochastic route choice model to get the total travel time. At last, we use the reverse of the total travel time as the fitness value.
4. **"Reach the maximum number of iterations?"**: this module checks first if the maximum number of generations has been reached; if not, proceed to the following steps.
5. **"Tournament"**: This module is used in order to obtain two parents from the last generation as a preparation for the next generation. In this module, we randomly select two chromosomes from the population, followed by a tournament between these two chromosomes and comparing their fitness function values. Higher valued chromosome won this tournament. At last, return the winner as one of the parents.
6. **"Crossover"**: Two chromosomes are selected using the "tournament" module, and the crossover happens under a pre-set probability (called probability of crossover. For each child, an inherent index $x_{inherent}$ is randomly selected as a float which is in the range of (0,1). Then the child's chromosome is calculated as in Equation (9).

$$Child = Father * x_{inherent} + Mother * (1 - x_{inherent}) \quad (9)$$

7. **"Mutation"**: Mutation changes the chromosome in children in a pre-set probability (called probability of mutation). In this application, mutation function only mutates between phases within one intersection. The reason is to maintain the cycle time in each intersection. For example, one child has a chromosome of:

$$[p_{11}, p_{12}, p_{13}, p_{14}, p_{21}, p_{22}, p_{23}, p_{24}, \dots, p_{n1}, p_{n2}, p_{n3}, p_{n4}].$$

We then randomly select: a) an intersection $u$ b) two phases $v$ and $w$ from this interection and c) the variation ($Var$) within the range of (0, $p_{uv}$). At last, the new duration of phases $v$ and $w$ are calculated as: $p'_{uv} = p_{uv} - Var$, $p'_{uw} = p_{uw} + Var$. The rest phase durations of this child remain the same.

8. **"GA optimization"**: continue to the next generation by going to step 2 until the stopping criteria has been reached.

**Table 1** Configuration of traffic signals for each intersection

| Phase ID | Traffic signal configuration (green movements highlighted) |
|---|---|
| 1 | 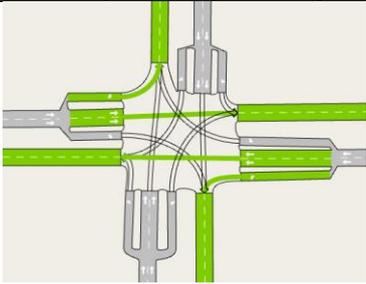 |
| 2 | 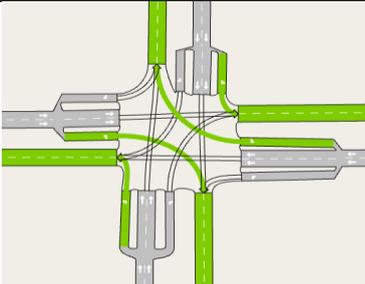 |
| 3 | 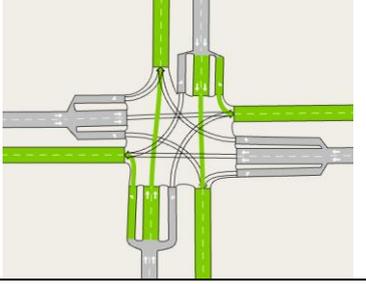 |
| 4 | 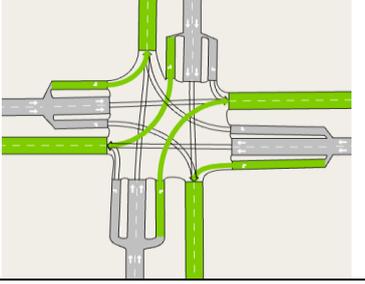 |

**Table 2** Traffic demand

| From\To | 1 | 2 | 3 | 4 | 5 | 6 | 7 | 8 | Total |
|---|---|---|---|---|---|---|---|---|---|
| 1 | 0 | 150 | 150 | 150 | 150 | 100 | 100 | 150 | 950 |
| 2 | 150 | 0 | 100 | 100 | 100 | 150 | 150 | 100 | 850 |
| 3 | 150 | 100 | 0 | 150 | 100 | 100 | 100 | 150 | 850 |
| 4 | 100 | 150 | 100 | 0 | 150 | 100 | 150 | 150 | 900 |
| 5 | 150 | 100 | 100 | 150 | 0 | 150 | 150 | 100 | 900 |
| 6 | 100 | 100 | 100 | 100 | 0 | 0 | 150 | 100 | 650 |
| 7 | 100 | 150 | **750** | 150 | 150 | 100 | 0 | 150 | **1550** |
| 8 | 100 | 150 | 150 | 100 | 150 | 100 | 100 | 0 | 850 |
| Total | 850 | 900 | **1450** | 900 | 800 | 800 | 900 | 900 | 7500 |

## 5. Case study

For showcasing the benefits of the proposed approach, a four-intersection network was designed in AIMSUN [22] and three scenarios are constructed in order to optimize the traffic signal timings under normal conditions and under traffic incident conditions. The GA model is tuned by running multiple times using different parameter settings before converging towards the optimal GA parameters to be used in the case study.

*5.1 Network configuration*

This network layout of the simulation model is shown in Fig. 2 (a) and is a left-hand drive model to accommodate the Australian road environment. The simulation duration is one hour and each intersection is a typical four-branch signalized intersection with dedicated right turning lane and dedicated left turn lane. The detailed layout of intersection #1 is shown in see Fig. 2 (b) as an example, and all the other intersections are configured in the same way.

This is an initial traffic network configuration for which we apply the proposed GA optimization. Further extension of this work will apply the methodology to a Sydney sub-network.

### 5.2 The configuration of traffic signals

Each intersection has the same cycle time (which is set to 90 seconds) and the same number of phases (which is 4). The order of phases is fixed. Within each phase, the green granted movements are the same and fixed for all intersections. The only variable in signal configuration is the phase green times. The configuration of traffic signals for each intersection is shown in Table 1.

### 5.3 Traffic demand

The O-D indexes are shown in Fig. 3 and the O-D trips for one-hour simulation are shown in Table 2. As highlighted in Table 2, a higher flow is set from centroid 7 to centroid 3. This O-D pair contains 2 routes, which are shown in Fig. 3 (Route 1 and Route 2). Special attention will be paid in observing and analysing the flows on these two routes under optimization constraint.

### 5.4 GA parameter tuning

Before applying the GA optimization method, there are several parameters that need to be set up which are: the population size, the maximum number of generations, the crossover probability, and the mutation probability, which have been tuned with the computational time in mind as well.

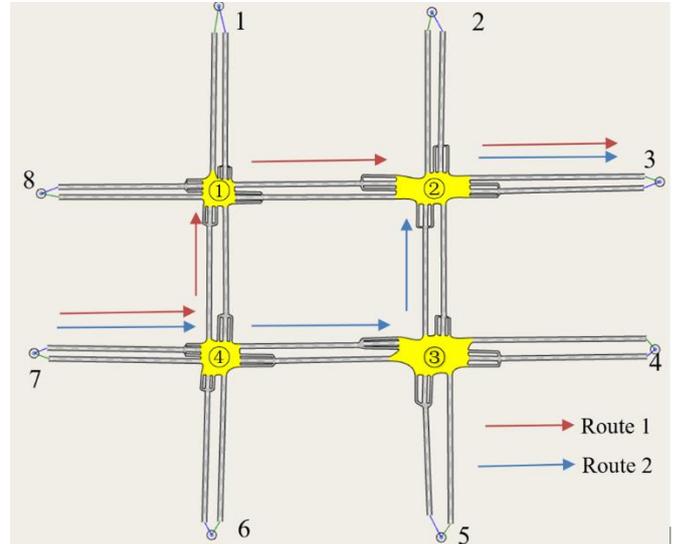

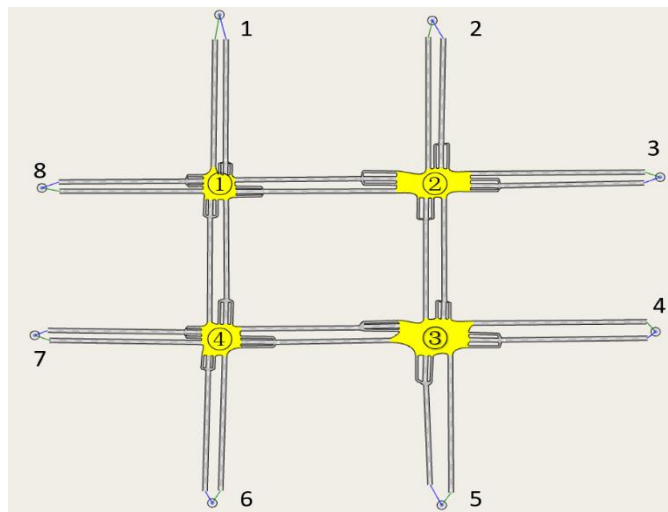

*a*

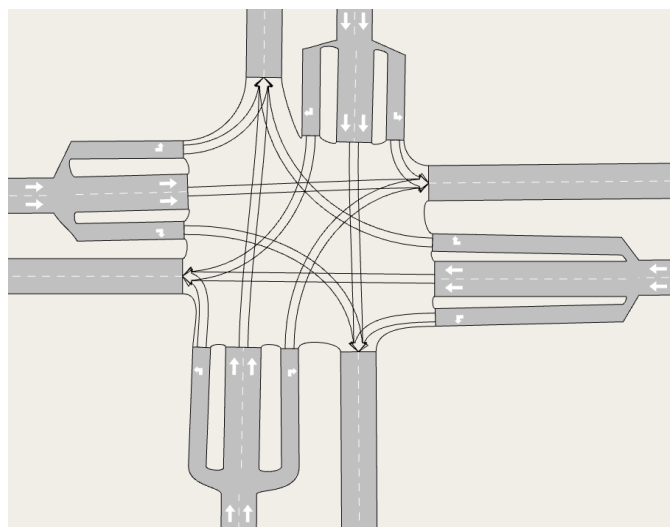

*b*

***Fig. 2.*** *(a)* Network layout, *(b)* Intersection #1 layout

***Fig. 3.*** *O-D index definition and major routes highlighted*

The details of these parameters are as follow:

- **Population size and maximum number of generations**: Population size is the number of individuals (chromosomes) in one population in one generation. Maximum number of generations is the maximum number of evolutional generations in one optimization. The max number of generations is determined by the performance of the fitness function and is set at the step after which the fitness function doesn't improve anymore. In addition, population size and maximum number of generations have direct relationship with total computational time which will be discussed in bullet "computational time" in the following paragraph.
- **Probability of crossover**: enables to inherit a good fitness from the last generation to new generation. This parameter must be very high in order to achieve fast convergence, so the probability of crossover is set to 0.8 in all experiments in this paper.
- **Probability of mutation**: Mutation generates new chromosomes which enrich the gene library. Mutation is a double-edged sword. On one hand, the mutation may happen to a chromosome with bad fitness and transform it into a chromosome with better fitness. On the other hand, mutation creates noise to the convergence of GA. In order to avoid noise in convergence, the mutation probability is set to 0.1 in all experiments of this paper.
- **Computational time**: Computational times are recorded at the beginning and the end of a generation. The most time-consuming procedure is the GA algorithm is the calculation of fitness value for each chromosome in each generation. Because the computation time for each fitness value calculation is relatively constant using Aimsun and the total number of fitness value calculations repetition is the product of population size and maximum number of generations, there is a linear relationship between accumulative computation time and the product of population size and maximum number of generations. Experiments show that the first 10 generations always consume more time than the rest of generations and after 10 generations, each generation takes the same time.

We tested the typical combination of GA parameters, then we determine a set of parameters with fast and stable convergence and relatively short computation time. The maximum number of iterations is set to 20, the population size is set to 75, the crossover probability is set to 0.8, and the mutation probability is set to be 0.1, which consumes an average computational time of about 7 minutes per generation.

Currently in our GA optimization process, we use sequential computing when calculating fitness values. In this experiment, the computational time for calculating one fitness value is less than 7 seconds. As we know, each individual in GA is independent and therefore, can be processed in parallel. Ideally, we can use parallel computing to physically reduce the computational time in the future. Ultimately, the computation time will not be the limitation of our algorithm.

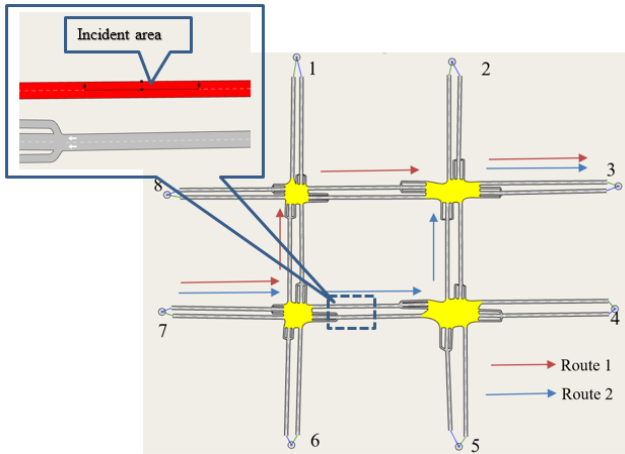

***Fig. 4.*** *Traffic incident configuration*

*5.5 Scenarios*

Using the above GA parameters, three scenarios are designed for our case study which are:
1. **No traffic incident scenario but using GA for traffic control optimization**: the proposed GA model will be applied to the "no-incident network" and a simulation applying the optimal signal control (we can call it "no-incident optimal signal control") to the "no-incident network" is recorded.
2. **Traffic incident scenario without GA traffic control optimization**: an incident is created in the network at the location shown in Fig. 4 which will last for one hour. The incident blocks one lane of a two-lane link in route 2 from centroid 7 to centroid 3. The traffic flows on both route 1 and route 2 will be affected by this incident. The traffic signal plan in scenario 2 is the same as scenario 1.
3. **Traffic incident scenario with the GA traffic control optimization**: the proposed GA model will be applied to the network and a simulation using the new optimal signal control will be recorded.

## 6. Results

*6.1 Scenario 1: No incident scenario with GA*

Let's denote $\{a_i, b_i, c_i, d_i, i = 1,..4\}$ as the phases of each intersection, where $a_1$ is the first phase of intersection 1, $b_1$ is the second phase of intersection 1, etc. The outcome of proposed GA model returned the following optimal phase values [in seconds] of the whole network under no incident conditions:

$$Optimal\ phase\ setting\ scenario\ 1 = \{18, 22, 12, 38, 20, 19, 15, 36, 17, 12, 17, 44, 30, 22, 9, 29\}$$

The corresponding optimal fitness value is -22.41, which corresponds to a total travel time of 22.41 $vehicle \cdot hour$.

The convergence of each phase in intersection 1 and intersection 3 towards the optimised solution are presented in Fig. 5 and 6 respectively. The convergence of each phase in other intersections has the same pattern as intersection 1 (see Annex A).

In each sub-figure (such as (a), (b), (c) and (d)), GA started with a big range of phase durations with scattered corresponding fitness values in generation 1. Then after various generations of evolution, the fitness values increases gradually and all phases have reached convergence at the end of the GA process in generation 20.

There is a significant trend for intersection 3 where the duration of phase 4 is getting longer as the number of generations increases. As shown in Table 1, phase 4 contains the right-turn movement of north and southbound traffic and left-turn movement from east and westbound traffic. The reason for this trend is the high demand from centroid 7 to centroid 3 shown in Table 2 and Fig. 3, which leads to high flows using route 1 and route 2 as shown in Fig. 3. The increasing trend in phase 4 duration in intersection 3 provides more green time to accommodate the traffic flows using route 2.

In addition, the simulated flow using the optimal traffic signal timings generated from GA model is presented in Fig. 7. The simulated flows prove that the optimal signal timings generated by the GA model are aware of the high demand and diverge the flows for two routes. The flows along route 1 and route 2 are around 1,200 to 1,300 vehicle/hr which are quite even. The reason for a evenly split between route 1 and route 2 flows is that both route 1 and route 2 has similar length, capacities, and turnings in our network.

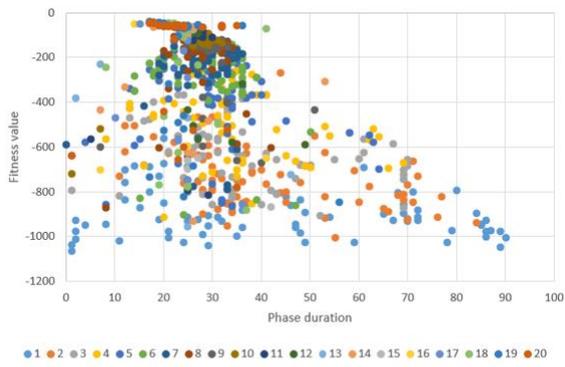
(a) Phase 1

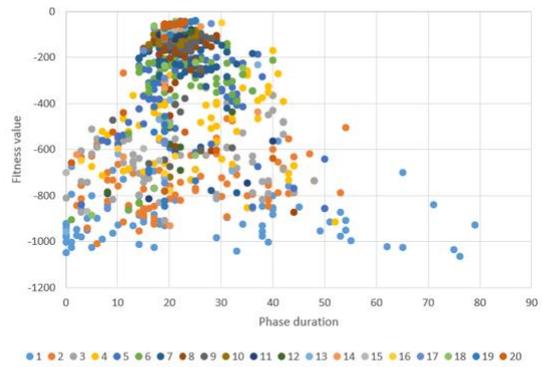
(b) Phase 2

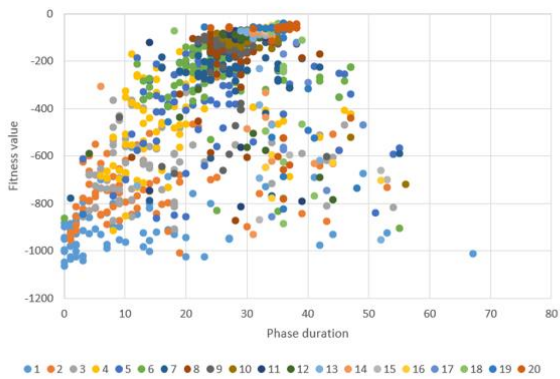
(c) Phase 3

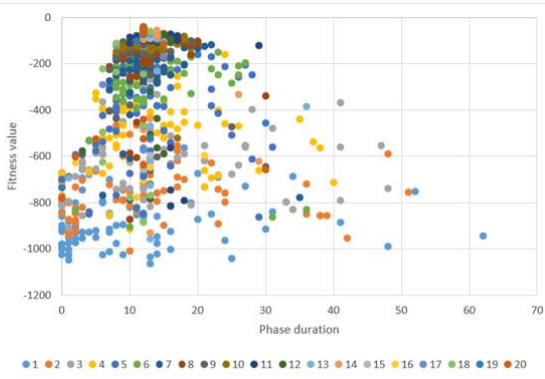
(d) Phase 4

*Fig. 5.* *Phase duration convergence in intersection 1*

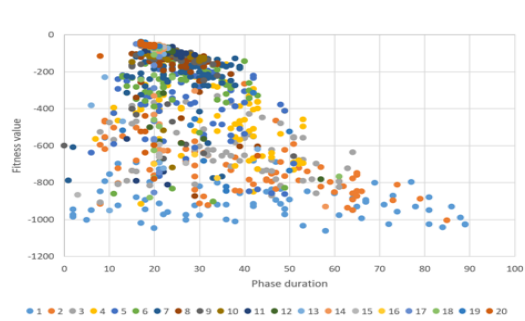
(a) Phase 1

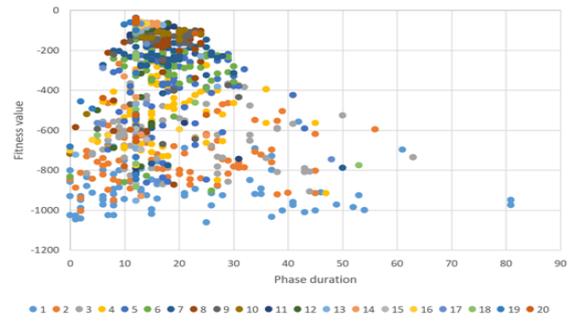
(b) Phase 2

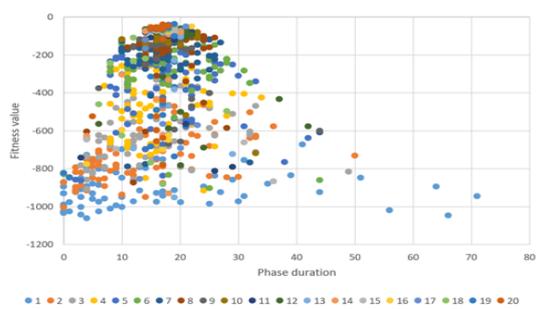
(c) Phase 3

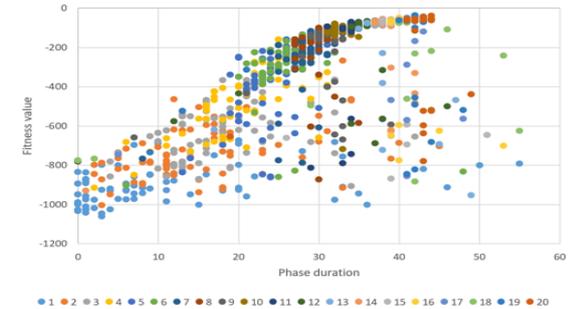
(d) Phase 4

*Fig. 6.* *Phase duration convergence in intersection 3*

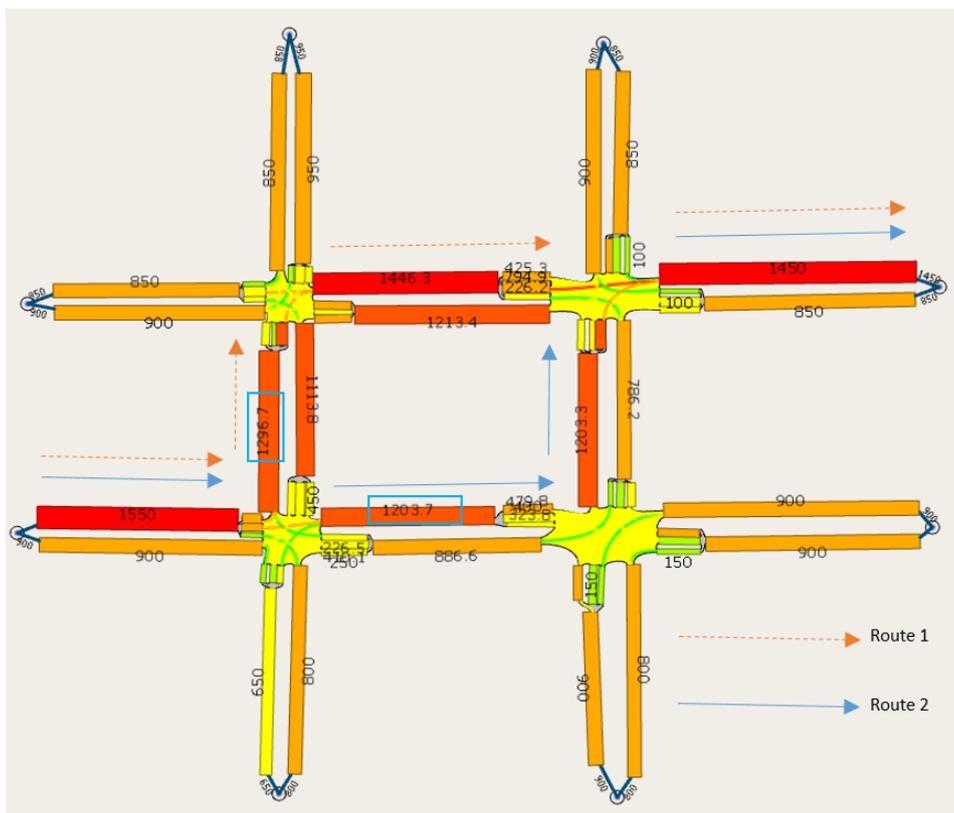

*Fig. 7.* Simulated flow under optimal traffic signal settings without any incident

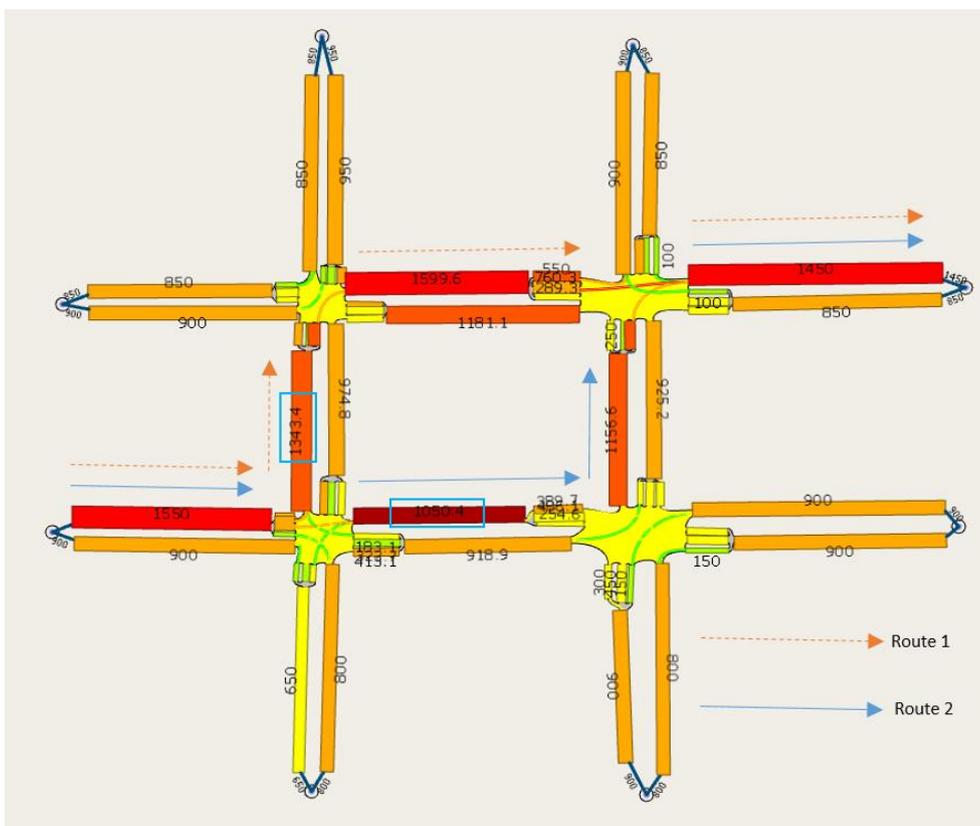

*Fig. 8.* Simulated flow with the incident

### 6.2 Scenario 2: Traffic incident scenario without GA

In this scenario, the same signal control plan as scenario 1 is used and the simulated flows are presented in Fig. 8. The total travel time obtained in this case is 47.37 $vehicle \cdot hour$, which is 111.38% more than the travel time experienced under no incident conditions (22.41 $vehicle \cdot hour$).

By comparing Fig. 7 and Fig. 8, the traffic flow on route 1 increased while the traffic flow on route 2 decreased. This is reasonable because there is an incident happening during the simulation on route 2 where drivers try to use route 1 instead of route 2. However, the traffic signals do not adapt to the shifting flows, therefore the flow in route 1 didn't increase too much.

### 6.3 Scenario 3: Traffic incident scenario with GA

In this scenario, the outcome of the proposed GA model is recorded. The convergence of each phase in each intersection has the same pattern as in Fig. 5 and can be found in Annex **B**. The final outcome of the GA model for this scenario is:

$Optimal\ phase\ setting\ scenario\ 3 =$
$\{31, 22, 13, 24, 29, 23, 17, 21, 30, 21, 18, 21, 29, 38, 9, 14\}$,

and the corresponding optimal fitness value is -28.24, which means total travel time is 28.24 $vehicle \cdot hour$ which is 26.02% more than the travel time experienced under no incident condition in scenario 1 (22.41 $vehicle \cdot hour$) but 40.76% lower than scenario 2 (47.37 $vehicle \cdot hour$). This means the GA model is capable of reducing the total travel time under non-recurrent incidents.

By comparing the phase durations between scenario 1 and 3, not all phases containing movements in route 2 are increased. For example, 4th phase in intersection 1 and 4th phase in intersection 4. On the contract, the other phases containing movements which are not affected by the incidents are given more green times more or less. Therefore, we can infer that one important source of total travel time saving comes from the sections that are not affected by the incident.

In addition, the simulated flow using the optimal traffic signal timings generated from GA model is presented in Fig. 9. The flow on the incident located section dropped comparing to Fig. 7. On the other hand, by comparing Fig. 8 and Fig. 9, the allocation of trips alone route 1 and route 2 are almost the same, which means that the GA optimized signals are adapted to the traffic flow.

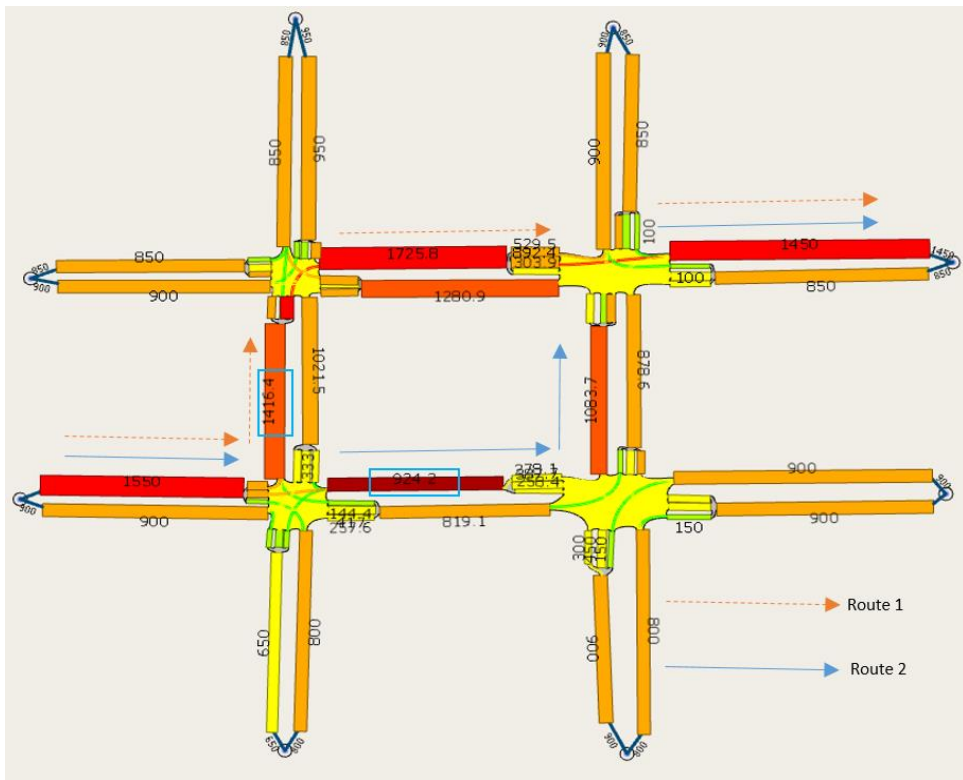

***Fig. 9.*** *Simulated flow under incident with GA optimized signal control*

### 6.4 Findings

In scenario 1, we simulated the daily normal traffic under normal traffic control plan. The GA model was applied to get the optimal traffic control plan. Then in scenario 2, a traffic incident was created in the network, and no more action was taken to response the traffic incident. The total travel time in scenario 2 increased by 111.38% compared to the total travel time in scenario 1. At last, we simulated the case that we took the instant response to the traffic incident and apply the GA model to re-estimate the optimal traffic control plan. The total travel time in scenario 3 only increased by 26.02% compared to the total travel time in scenario 1.

By comparing the outputs of scenario 2 and scenario 3, we conclude that the proposed GA model is able to adjust the signal timings to minimize the total travel time. In our case study, 40.76% of total travel time saving is achieved in our network.

## 7. Conclusion

In this paper, a GA method is developed to mitigate the impact of non-recurrent traffic incidents. A four-intersection network is designed as the experiment base model in AIMSUN. The proposed GA model is transformed from a standard GA model by adapting the key components to traffic signal timing optimization. These components consist of initialization, fitness function calculation, crossover, mutation and so on. A proper set of GA parameters are chosen according to the prior experiments in order to achieve fast and stable convergence and short computational time. At last, the experiment is designed to simulate the cases whether TMC takes action to revise traffic control plans after the appearance of an incident or not. The experiment results show improvement of total travel time if the TMC uses the proposed GA model to re-optimize the traffic control plan under the incident condition comparing to taking no action at all. The saving in total travel time is 40.76%.

Future work can be done in investigation more complicated network and even real-world network. In order to fit the proposed model to real-world application, work can be done in parallel computing to further shorten the computation time and further increase the efficiency. In addition, apply reinforcement learning to further speed up the convergence speed in GA is also a good perspective.

## 8. Acknowledgement

The work presented in this paper is funded by the New South Wales Premiere's Innovation Initiative. The authors of this work are grateful for the work and support of the Transport for New South Wales, Australia. Data61 is funded by the Australian Federal Government through the Commonwealth Scientific and Industrial Research Organization.

## 9. Author contribution

The authors confirm contribution to the paper as follows: study conception, design, and validation: Dr. Mao and Dr. Mihaita; data science and model performance: Dr. Mao; draft manuscript preparation: all authors. All authors reviewed the results and approved the final version of the manuscript.

## 10. Reference

[1] B. Anbaroglu, B. Heydecker, and T. Cheng, "Spatio-temporal clustering for non-recurrent traffic congestion detection on urban road networks," Transportation Research Part C: Emerging Technologies, vol. 48, pp. 47-65, 2014.
[2] A. Skabardonis, P. Varaiya, and K. Petty, "Measuring recurrent and nonrecurrent traffic congestion," Transportation Research Record: Journal of the Transportation Research Board, no. 1856, pp. 118-124, 2003.
[3] P. P. Varaiya, Finding and Analyzing True Effect of Non-recurrent Congestion on Mobility and Safety. California PATH Program, Institute of Transportation Studies, University of California at Berkeley, 2007.
[4] X. Ban, J. M. Wojtowicz, and W. Li, "Decision-making tool for applying adaptive traffic control systems," New York State Energy Research and Development Authority2016.
[5] P. Farradyne, "Traffic incident management handbook," Prepared for Federal Highway Administration, Office of Travel Management, 2000.
[6] B. Mehran, "Evaluation of possible directions for improving traffic management system," in Proceedings of the 93rd TRB annual meeting. Washington, DC [CD]. Retrieved from http://docs.trb.org/prp/13-1073.pdf, 2013.
[7] P. Nitsche et al., "Pro-active Management of Traffic Incidents Using Novel Technologies," Transportation Research Procedia, vol. 14, pp. 3360-3369, 2016.
[8] M. Haselkorn, S. Yancey, and S. Savelli, "Coordinated Traffic Incident and Congestion Management (TIM-CM): Mitigating Regional Impacts of Major Traffic Incidents in the Seattle I-5 Corridor," 2018.
[9] P. Charles, Traffic incident management: best practice (no. AP-R304/07). 2007.
[10] H. Nguyen, C. Cai, and F. Chen, "Automatic classification of traffic incident's severity using machine learning approaches," IET Intelligent Transport Systems, vol. 11, no. 10, pp. 615-623, 2017.
[11] T. Wen, A.-S. Mihăiță, H. Nguyen, C. Cai, and F. Chen, "Integrated Incident Decision-Support using Traffic Simulation and Data-Driven Models," Transportation Research Record, p. 0361198118782270, 2018.
[12] M. Papageorgiou, C. Diakaki, V. Dinopoulou, A. Kotsialos, and Y. Wang, "Review of road traffic control strategies," Proceedings of the IEEE, vol. 91, no. 12, pp. 2043-2067, 2003.
[13] C. Diakaki et al., "Extensions and new applications of the traffic-responsive urban control strategy: Coordinated signal control for urban networks," Transportation Research Record: Journal of the Transportation Research Board, no. 1856, pp. 202-211, 2003.
[14] S. G. Ritchie, "A knowledge-based decision support architecture for advanced traffic management," Transportation Research Part A: General, vol. 24, no. 1, pp. 27-37, 1990.
[15] D. E. Goldberg and J. H. Holland, "Genetic algorithms and machine learning," Machine learning, vol. 3, no. 2, pp. 95-99, 1988.
[16] M. D. Foy, R. F. Benekohal, and D. E. Goldberg, "Signal timing determination using genetic algorithms," Transportation Research Record, no. 1365, p. 108, 1992.
[17] H. Ceylan and M. G. Bell, "Traffic signal timing optimisation based on genetic algorithm approach, including drivers' routing," Transportation Research Part B: Methodological, vol. 38, no. 4, pp. 329-342, 2004.
[18] A. S. Mihăiță, L. Dupont, and M. Camargo, "Multi-objective traffic signal optimization using 3D mesoscopic simulation and evolutionary algorithms," Simulation Modelling Practice and Theory, vol. 86, pp. 120-138, 2018.
[19] B. Pan, U. Demiryurek, C. Gupta, and C. Shahabi, "Forecasting spatiotemporal impact of traffic incidents for next-generation navigation systems," Knowledge and Information Systems, vol. 45, no. 1, pp. 75-104, 2014.
[20] H. Yang and S. Yagar, "Traffic assignment and signal control in saturated road networks," Transportation Research Part A: Policy and Practice, vol. 29, no. 2, pp. 125-139, 1995/03/01/ 1995.
[21] M. Smith and T. Van Vuren, "Traffic equilibrium with responsive traffic control," Transportation science, vol. 27, no. 2, pp. 118-132, 1993.

[22] T. Aimsun, "Dynamic simulators users manual," Transport Simulation Systems, vol. 20, 2012.

## 11. Annex A

Scenario 2: Traffic incident scenario without GA

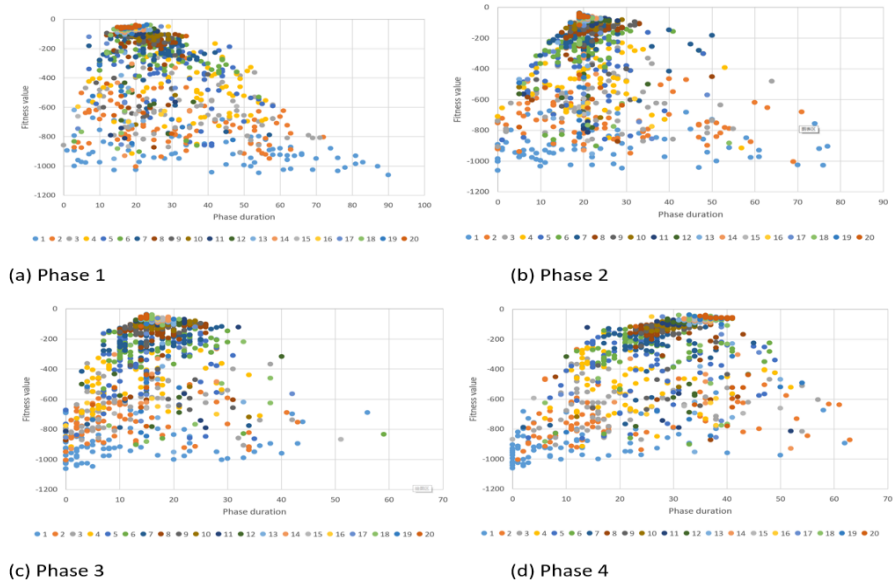

*Fig. 10. Phase duration convergence in intersection 2*

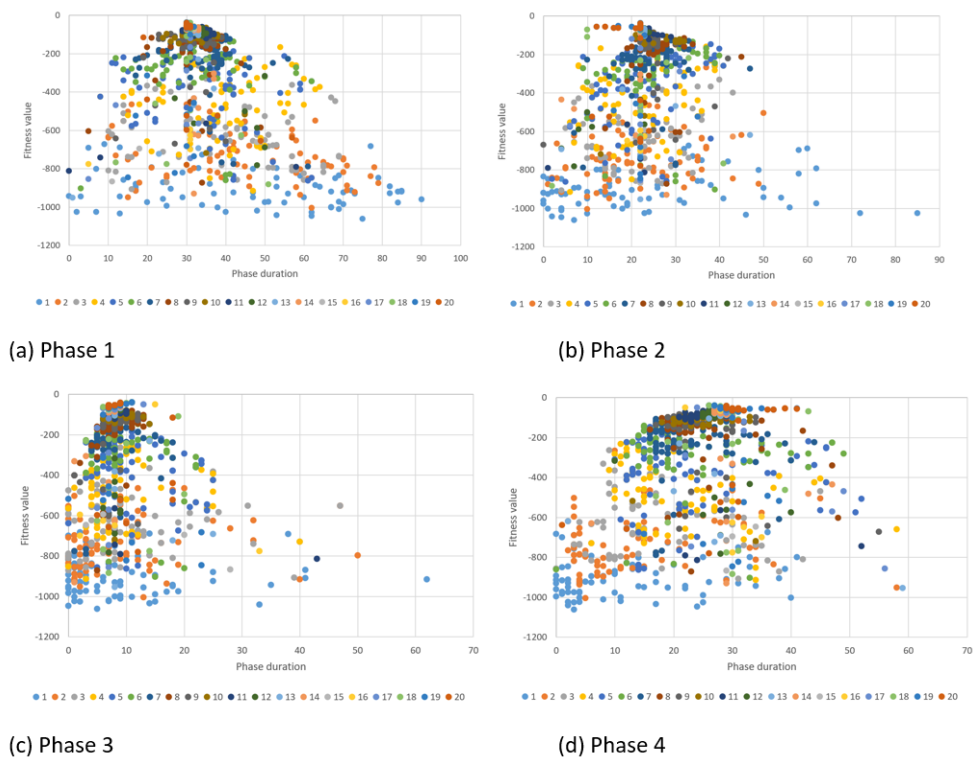

*Fig. 11. Phase duration convergence in intersection 4*

## 12. Annex B

Scenario 3: Traffic incident scenario with GA

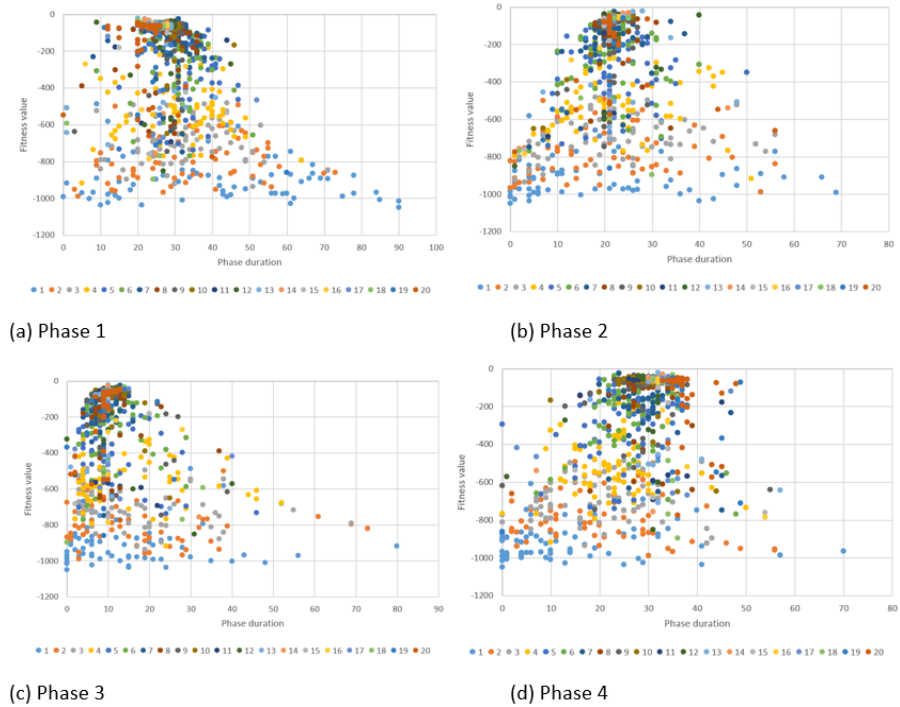

*Fig. 12. Phase duration convergence in intersection 1*

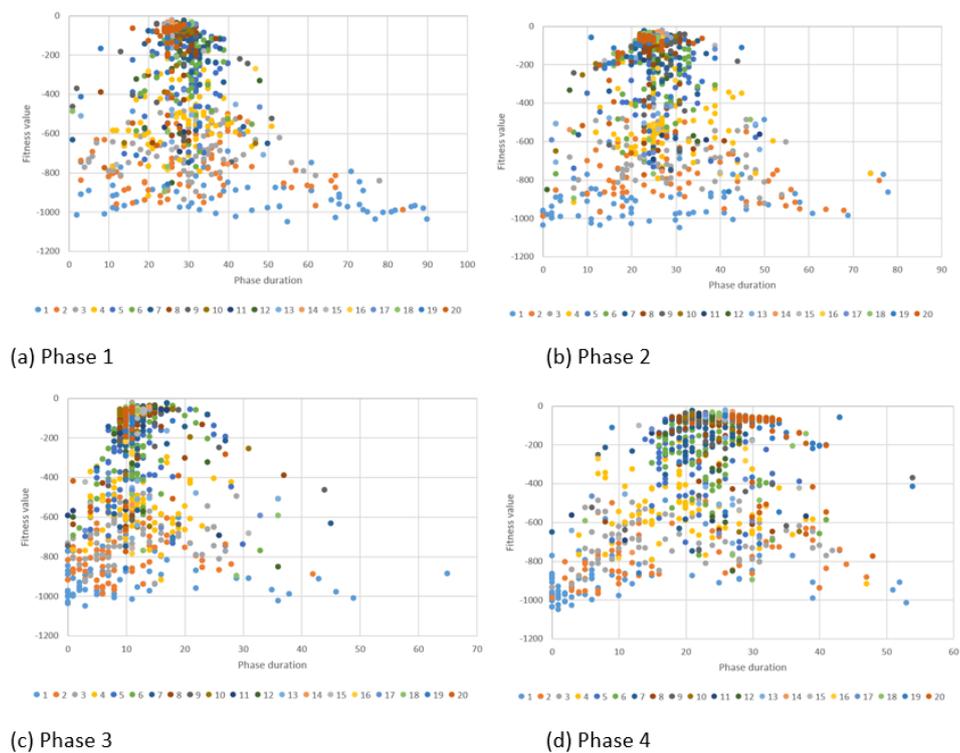

*Fig. 13. Phase duration convergence in intersection 2*

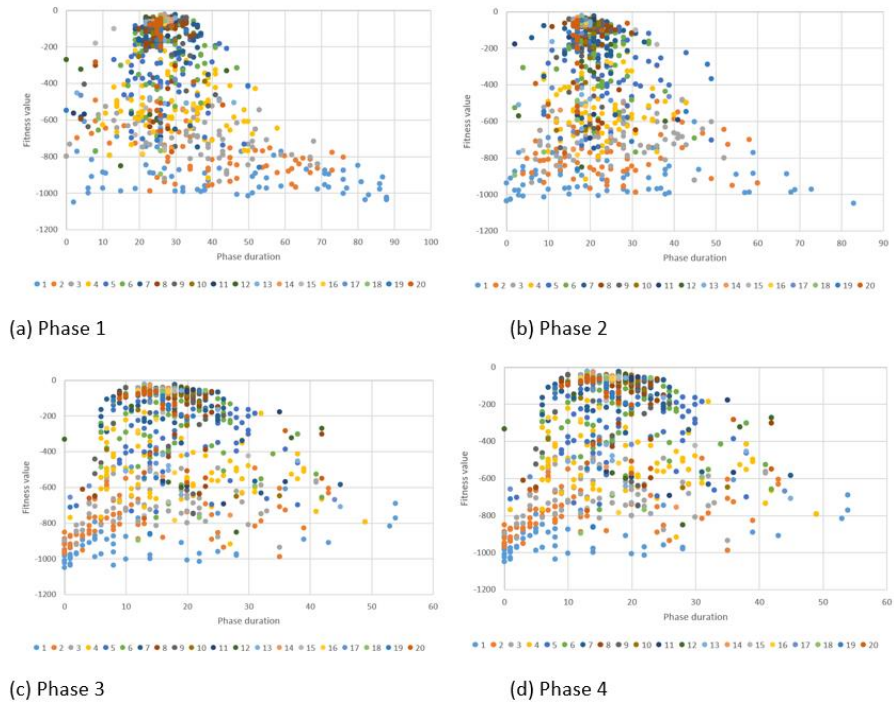

*Fig. 14. Phase duration convergence in intersection 3*

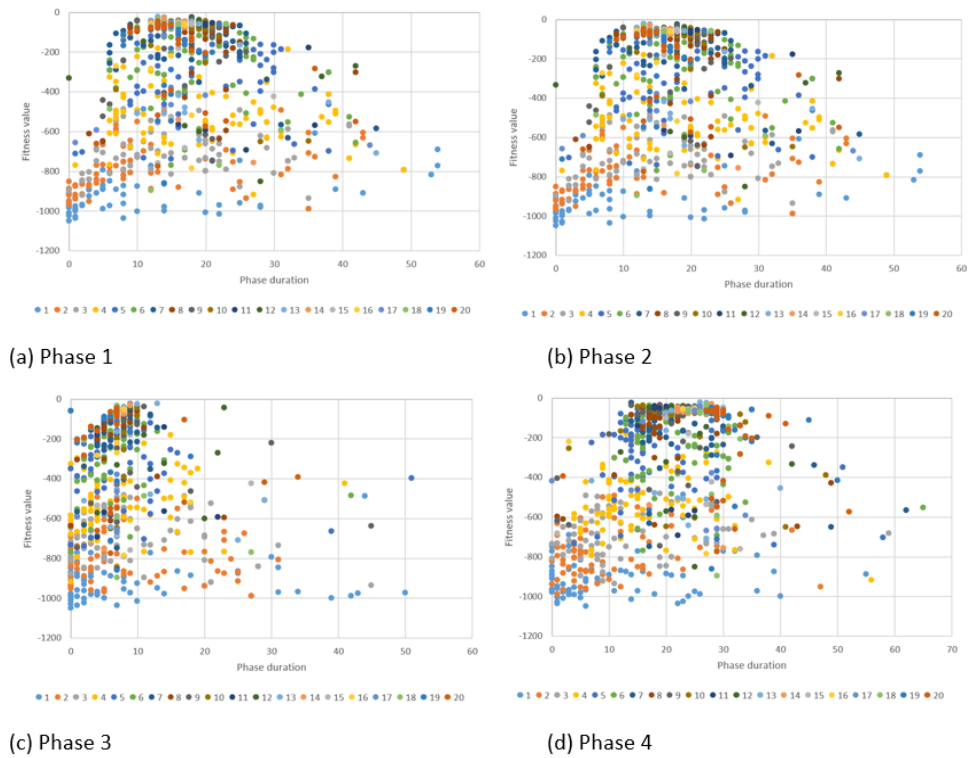

*Fig. 15. Phase duration convergence in intersection 4*